\documentclass[%
 reprint,
 amsmath,amssymb,
 aps,
]{revtex4-1}
\usepackage{graphicx}
\usepackage{dcolumn}
\usepackage{bm}
\usepackage{comment}
\usepackage{upgreek}

\begin{document}

\newcommand{\yimp}{$y_\mathrm{imp}$ }
\newcommand{\ymelt}{$y_\mathrm{melt}$ }
\newcommand{\linny}{$\gamma_\mathrm{L}$ }
\newcommand{\ymeltL}{$\langle y_\mathrm{melt} \rangle/L$ }
\newcommand{\yimpL}{$y_\mathrm{imp}/L$ }

\preprint{APS/123-QED}

\title{Local melting attracts grain boundaries in colloidal polycrystals}

\author{Caitlin E. Cash}%
 \affiliation{%
Department of Physics,  Harvey Mudd College, Claremont, CA
}%

\author{Jeremy Wang}%
 \affiliation{%
 Department of Physics, Harvey Mudd College, Claremont, CA
}%

\author{Maya M. Martirossyan}%
 \affiliation{%
 Department of Physics, Harvey Mudd College, Claremont, CA
}%

\author{B. Kemper Ludlow}%
 \affiliation{%
 Department of Physics, Harvey Mudd College, Claremont, CA
}%

\author{Alejandro E. Baptista}%
 \affiliation{%
 Department of Physics, Harvey Mudd College, Claremont, CA
}%

\author{Nina M. Brown}%
 \affiliation{%
 Department of Physics, Harvey Mudd College, Claremont, CA
}%

\author{Eli J. Weissler}%
 \affiliation{%
 Department of Physics, Harvey Mudd College, Claremont, CA
}%

\author{Jatin Abacousnac}%
 \affiliation{%
 Department of Physics, Harvey Mudd College, Claremont, CA
}%

\author{Sharon J. Gerbode}
 \email{gerbode@hmc.edu}
\affiliation{%
 Department of Physics, Harvey Mudd College, Claremont, CA
}%

\date{\today}

\begin{abstract}
We find that laser-induced local melting attracts and deforms grain boundaries in 2D colloidal crystals. When a melted region in contact with the edge of a crystal grain recrystallizes, it deforms the grain boundary --- this attraction is driven by the multiplicity of deformed grain boundary configurations. Furthermore, the attraction provides a method to fabricate artificial colloidal crystal grains of arbitrary shape, enabling new experimental studies of grain boundary dynamics and ultimately hinting at a novel approach for fabricating materials with designer microstructures.
\end{abstract}

\pacs{'82.70.Dd'}

\maketitle

\begin{figure}[h!]
 \includegraphics[width=0.48\textwidth]{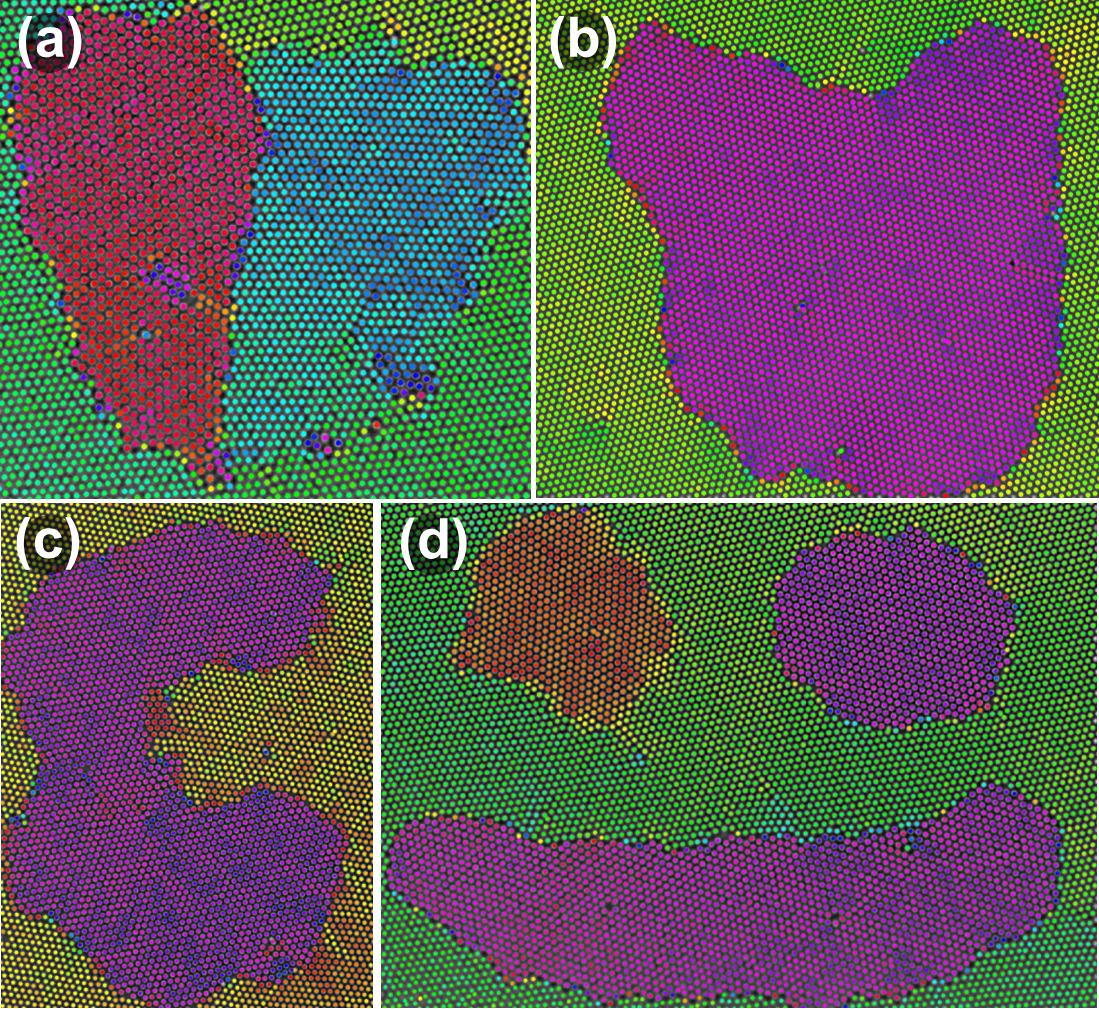}
  \caption{Experimentally fabricated grains. Local melting allows the creation of grains with arbitrary shapes: (a) ``broken heart'' (b) ``cat'' (c) ``C'' (d) ``smiley face.'' Particles are colored by the phase of the local orientational order parameter $\Psi_6$, indicating grain orientation.}
  \label{fig:Zoo}
\end{figure}

Grain boundaries, the disordered interfaces between neighboring crystal grains, play a vital role in phase changes such as melting \cite{Chui1983,Li2016}, and determine important material properties ranging from yield strength to electrical conductivity \cite{Meyers2006,Kreuer2003,Huang2011}. While measurements of the atom-scale dynamics of grain boundaries are limited in solid state crystals, experimental studies of grain boundaries in colloidal crystals have offered a particle-scale view of thermal grain boundary fluctuations \cite{Skinner2010,Gokhale2013} and responses to macroscopic mechanical perturbation \cite{Wei2004,Buttinoni2017}. Furthermore, self-assembled crystals of colloidal particles are interesting in their own right; such systems promise to be a useful avenue for fabricating microstructured materials such as photonic crystals \cite{Kim2011}. The presence of grain boundaries in such colloidal materials can influence performance, so the ability to rearrange individual grain boundaries could allow for detailed tuning of materials properties. For example, a layered grain structure could create an elastically anisotropic material, with a compliant axis and a stiff axis, reminiscent of behavior in multilayered thin film crystals \cite{Armstrong2013,Armstrong1965}. Here we report experiments in which a focused laser beam causes local melting within colloidal polycrystals. We find that local melting attracts grain boundaries, allowing grain boundary manipulation and enabling the fabrication of arbitrarily shaped grains (Fig.~\ref{fig:Zoo}). We systematically study the attraction by conducting a simple experiment in which a grain boundary is locally deformed, and use a lattice model to explain the resulting deformation. Ultimately, our discovery hints at a new approach for engineering materials using internal grain manipulation to create designer microstructures.

We prepare colloidal suspensions of silica spheres, diameter $1.2~\upmu$m (Sekisui Micropearl Spacers, Dana Enterprises International, CA),  in dimethyl sulfoxide. The suspension is pipetted into a wedge-shaped cell constructed from two glass coverslides as previously described \cite{Gerbode2008}. The cell is tilted to allow the particles to sediment into the gap between the coverslides, where they form an effectively 2D crystalline monolayer. At visible wavelengths, the refractive index  of the silica spheres is approximately 1.424, while that of dimethyl sulfoxide is 1.477. Because the refractive index of the particles is \emph{less} than that of the fluid, a standard optical tweezer setup (800~mW, 1064~nm laser focused through a 100X Zeiss Plan-Apochromat objective) yields repulsive forces on the particles -- ``optical blasting'' rather than optical trapping.

\begin{figure}[h!]
  \includegraphics[width=0.48\textwidth]{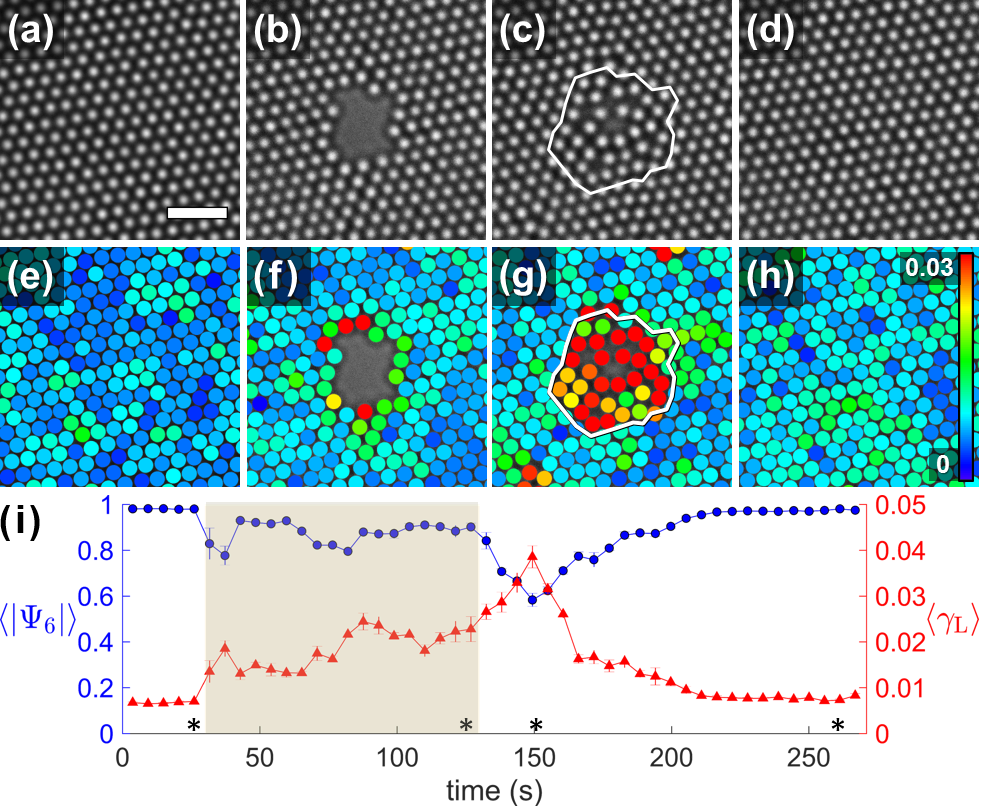}
  \caption{Optical blasting causes local melting in a 2D colloidal crystal (scale bar: 5~$\upmu$m). A crystal is shown (a) before blasting, (b) during the blast, (c) melted after the blast, and (d) recrystallized. The melted region is outlined in white. The same region is shown in (e) - (h), with particles colored by modified dynamic Lindemann parameter $\gamma_\mathrm{L}$, which quantifies local fluctuations in particle separations (color bar in (h) applies to (e) -- (h)). (i) Time series of the average values of the magnitude of the orientational order parameter $|\Psi_6|$ (blue dots) and \linny  (red triangles), measured within the white outline over the entire experiment (error bars: SEM).  The shaded region denotes when the laser is on, and the asterisks indicate the frames shown in (a) -- (d).}
  \label{fig:Blast1}
\end{figure}

Optical blasting within a 2D colloidal crystal displaces particles within the plane radially outward from the laser focus, creating a hole (Figs.~\ref{fig:Blast1}(a) and \ref{fig:Blast1}(b)). When the laser is turned off, particles diffuse back into the hole to form a locally melted region (white outline in Fig.~\ref{fig:Blast1}(c)), which ultimately recrystallizes (Fig.~\ref{fig:Blast1}(d)). Preliminary experiments reveal the same behavior in 3D colloidal crystals, though confocal microscopy is needed to further study this effect.

\begin{figure*}[ht!]
  \includegraphics[width=\textwidth]{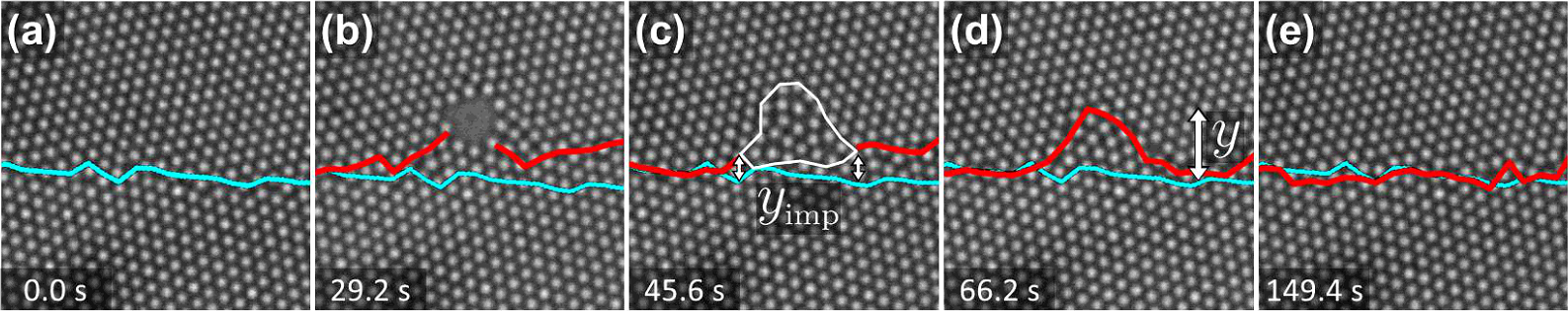}
  \caption{Optical blasting attracts grain boundaries. (a) An initially flat grain boundary (cyan). (b) An optical blast disrupts the grain boundary (red) from its original position (cyan). (c) When a melted region (white outline) forms after the blast, the grain boundary has been deformed by a distance \yimp relative to the original grain boundary position. The distance \yimp  is the average of the displacements on the left and right edges of the melted region. (d) Recrystallization pulls the grain boundary toward the blast location even further, for a total displacement $y$ above the original grain boundary. (e) Finally, the grain boundary relaxes down to a flat configuration.}
  \label{fig:BlastGB}
\end{figure*}

We determine the degree of disorder of the particles within the melted region using the local orientational order parameter $\Psi_6 = \frac{1}{N}\sum_n e^{6i\Theta_n}$, where the sum is taken over a particle's $N$ nearest neighbors (determined using Delaunay triangulation) and $\Theta_n$ is the angle between the $n^{th}$~neighbor separation vector and the horizontal axis \cite{Halperin1978}. To avoid associating neighbors across the hole, we filter out neighbors with separations greater than $1.6$ lattice constants (LC), corresponding to about $2.25~\upmu$m. As shown in Fig.~\ref{fig:Blast1}(i), the average value of $|\Psi_6|$ within the white outline decreases slightly when the laser is switched on. In this study, we are most interested in what happens once the laser is off: $\langle |\Psi_6| \rangle$ sharply drops as the melted region forms, and finally increases again as the region recrystallizes.

To quantify the fluidity of the particles in the melted region, we use a modified dynamic Lindemann parameter \linny that measures the fluctuations in a particle's average nearest neighbor separation, relative to the lattice constant $d$: $\gamma_\mathrm{L} = \frac{\sqrt{ \langle [u(t)-\langle u(t)\rangle]^2 \rangle}}{d}$. Here  $u(t)$ is the average distance between the particle and its nearest neighbors at time $t$, and the angled brackets denote time average over a time interval $T = 1.4$~s (7 frames). Again, only neighbors within 1.6 LC are considered, to avoid neighbors across the hole. In Figs.~\ref{fig:Blast1}(e-h), the particles are overlaid with dots colored by the value of $\gamma_\mathrm{L}$. We determine the extent of the melted region by finding the 360-sided polygon centered at the blast location that encloses all particles with \linny greater than one standard deviation above the average value of \linny calculated throughout the entire frame. We define the lengthscale $L$ of the melted region as the square root of the area of this polygon. The average value of \linny for particles within the white outline increases sharply when the melted region forms and decreases during recrystallization (Fig.~\ref{fig:Blast1}(i)). 

To understand the mechanism that enables us to create artificial grains (Fig.~\ref{fig:Zoo}), we investigated the effect of isolated local melting near a grain boundary. Indeed, we find that optical blasting near a grain boundary deforms the grain boundary toward the blast (Fig.~\ref{fig:BlastGB}). We explore this effect by conducting 94 optical blasting experiments, each with blast duration of $30$~s and blast location approximately $4$ LC above a flat grain boundary, which maximizes the size of the deformation in the experiments. To quantify how the deformation size depends on blast location, we conducted Brownian Dynamics simulations, in which the deformation size is also maximized at $4$ LC (see the Supplemental Material). We calculate the grain boundary position in each frame using a technique similar to that described in \cite{Skinner2010}, in which the image is divided into vertical bins and the grain boundary position in each bin is determined to be the height at which the crystal orientation switches from that of the upper grain to that of the lower grain. A detailed description of this method is provided in the Supplemental Material. 

We find that first, when the laser is switched on, a hole forms that disrupts the grain boundary and shifts it upward as shown in Figs.~\ref{fig:BlastGB}(a) and \ref{fig:BlastGB}(b). Then, when the laser is switched off, particles flow back into the hole. Once all flow has ceased, a locally melted region forms (white outline in Fig.~\ref{fig:BlastGB}(c)) and subsequently recrystallizes, pulling the grain boundary further upward so that the lower grain protrudes into the upper grain (Fig.~\ref{fig:BlastGB}(d)). This perturbation eventually relaxes, and the grain boundary returns to a flat configuration after about 1.5~minutes (Fig.~\ref{fig:BlastGB}(e)).
 
The deformation of the grain boundary proceeds through two distinct attraction mechanisms. The first occurs while the laser is on, and the grain boundary is attracted upward to the hole, which acts like an immobile impurity (see Supplemental Material for a detailed analysis). This attraction is reminiscent of previously reported interactions between grain boundaries and large impurities in colloidal crystals \cite{Villeneuve2005,Villeneuve2009} and is not the main focus of this manuscript. When the laser is turned off and particles fill the hole, the grain boundary relaxes down slightly, reaching the bottom of the melted region for a net upward displacement of $y_\mathrm{imp}$ due to this ``impurity effect'' (Fig.~\ref{fig:BlastGB}(c)). We find that on average over the 94 experiments, $y_\mathrm{imp} = 1.3 \pm 0.1$ LC (mean~$\pm$~SEM).

After the laser is turned off and particles have refilled the hole, the resulting melted region pulls the grain boundary even further upward. This second attraction mechanism is the primary focus of this paper as it provides the ability to move grain boundaries and tailor grain size and shape. On average, as the melted region recrystallizes, the grain boundary rises by an additional $2.2 \pm 0.2$ LC, for a total  displacement of $y$ measured to the peak of the protrusion, relative to the initial grain boundary position (Fig.~\ref{fig:BlastGB}(d)). 

\begin{figure}[t!]
  \includegraphics[width=0.48\textwidth]{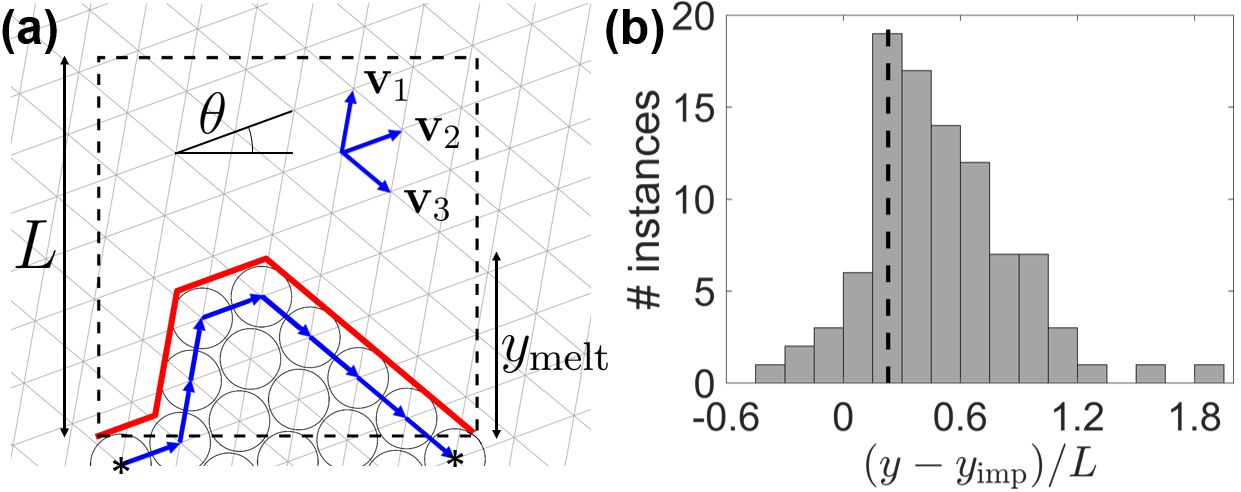}
  \caption{Lattice model explains attraction to melted region. (a) Schematic diagram of lattice model. A portion of the lower grain (circles) extends into the melted region (dashed square) by a distance $y_\mathrm{melt}$. A single grain boundary segment (red) corresponds to a particular sequence of lattice vectors (blue arrows) between adjacent particles at the top of the grain.  Each step in the sequence follows one of the three lattice vectors $\mathbf{v}_1$, $\mathbf{v}_2$, $\mathbf{v}_3$. The angle $\theta$ indicates the orientation of the lattice relative to the bottom edge of the melted region, which has side length $L$. (b) Distribution of vertical grain boundary displacements as a fraction of the size of the melted region. The  theoretical fractional distance $\langle y_\mathrm{melt} \rangle/L$ predicted by the lattice model (vertical dashed line) is consistent with the experimental fractional distances $\left(y - y_\mathrm{imp}\right)/L$ measured from 94 optical blasting experiments.}
  \label{fig:ToyModel} 
\end{figure}

This attraction of the grain boundary into the melted region seems counterintuitive: one might expect the melted region to recrystallize back into the surrounding upper crystal's orientation. To understand why the grain boundary shifts up, we use a lattice model to describe the recrystallization of the melted region and ultimately predict how far the lower crystal will protrude into the previously melted region. We model the melted region as a square of side length $L$ and find all possible grain boundary segments that remain within the square and are bounded at the bottom corners. To define a single grain boundary segment with peak height $y_\mathrm{melt}$, we place particles (circles in Fig.~\ref{fig:ToyModel}(a)) on lattice sites corresponding to the lower grain's orientation. The lattice vectors are $\mathbf{v}_1$, $\mathbf{v}_2$, and $\mathbf{v}_3$, with $\mathbf{v}_2$ oriented at an angle $\theta$ relative to the bottom edge of the melted region (Fig.~\ref{fig:ToyModel}(a)). 

The top layer of particles, adjacent to the grain boundary segment (red, Fig.~\ref{fig:ToyModel}(a)), can be represented by a sequence of lattice vectors (blue arrows, Fig.~\ref{fig:ToyModel}(a)). These vectors form a path corresponding to a grain boundary segment that starts at the bottom left corner of the melted region and ends at the bottom right corner. The path of vectors is bounded by the lattice sites below these corners (asterisks, Fig.~\ref{fig:ToyModel}(a)). Because the vectors represent a path between particle centers, while the grain boundary segment lies along particle edges, we approximate the arclength $s$ of a grain boundary segment (in LC) as the number of lattice vectors, plus one.

The Read-Shockley formula for the energetic cost per unit arclength of a 2D grain boundary is $\gamma_0 \phi(A - \ln \phi)$, where $\phi$ is the misorientation angle between the two grains, $\gamma_0$ has dimensions of energy per length, and both $A$ and $\gamma_0$ depend on $\theta$ and the elastic moduli \cite{Read1950}. We define $s_0 = k_BT /[\gamma_0 \phi(A - \ln \phi)]$ so that the energy of a grain boundary of arclength $s$ may be written simply as $E(s) = \frac{s}{s_0} k_BT$. We take each possible grain boundary segment to be a unique microstate $m$ of the system, with arclength $s_m$ and peak height  $y_{\mathrm{melt},m}$. The partition function is computed as
\begin{align}
	Z
    =
    \sum_m e^{-E(s_m)/k_BT}
    =
    \sum_m e^{-s_m/s_0} .
\end{align}

\noindent Using this model, the expected value of \ymelt is

\begin{align}
\langle y_\mathrm{melt} \rangle  =  \frac{1}{Z}\sum_m (y_{\mathrm{melt},m})  e^{-s_m/s_0}.
\end{align}

\noindent We consider the $\theta = 0$ case to provide intuition: at zero temperature, $\langle y_\mathrm{melt} \rangle$ would equal zero, because a straight grain boundary segment is the lowest energy configuration. However, at finite temperature, $\langle y_\mathrm{melt} \rangle > 0$ because there are more deformed than straight grain boundary configurations. 

We measure the experimental values of the parameters $s_0$ and $L$ to directly compare the model prediction with experimental data. To determine the value of $s_0$ from our experiments, we collect images of flat, unperturbed grain boundaries undergoing thermal fluctuations. The probability of a single grain boundary point fluctuation of height $h$ is proportional to the Boltzmann factor $e^{-s(h)/s_0}$, where $s(h)$ is the arclength added by the vertical fluctuation. The distribution of fluctuation heights can therefore be inverted to obtain the value of $s_0$. Using more than 250,000 individual grain boundary fluctuation measurements, we find that $s_0 = 0.33 \pm 0.03$ LC for our experiments. See Supplemental Material for a description of these experiments and their analysis.

We measure the average lengthscale $L$ in the optical blasting experiments to be $4.6 \pm 0.1$ LC. To compare experiments with different $L$ values and relate them to model predictions, we divide $\langle y_\mathrm{melt} \rangle$ by $L$. In the model, we find that $\langle y_\mathrm{melt} \rangle / L$ is not sensitive to the value of $L$ over the range from $L = 2$ to $L = 6$ (details in Supplemental Material). Using the experimental values for $s_0$ and $L$, we compute \ymeltL for values of $\theta$ spaced at $1^\circ$ intervals over the entire range from $-30^\circ$ to $30^\circ$, finding that the mean value of \ymeltL is $\langle y_\mathrm{melt} \rangle / L = 0.22$ (vertical dashed line in Fig.~\ref{fig:ToyModel}(b)).

To isolate the effect of the attraction mechanism that occurs after the melted region has formed, we plot the distribution of  $\left(y - y_\mathrm{imp}\right)/L$ measured from the optical blasting experiments (gray bars, Fig.~\ref{fig:ToyModel}(b)). The distance $\left(y - y_\mathrm{imp}\right)$ is equivalent to $y_\mathrm{melt}$; it is the displacement into the melted region caused by the second attraction. As shown in Fig.~\ref{fig:ToyModel}(b), the predicted value $\langle y_\mathrm{melt} \rangle/L$ is consistent with the measured values of $\left(y - y_\mathrm{imp}\right)/L$, indicating that the attraction to the melted region is driven by the greater multiplicity of curved grain boundary segments, despite their extra energetic cost.

\begin{figure}[t!]
  \includegraphics[width=0.48\textwidth]{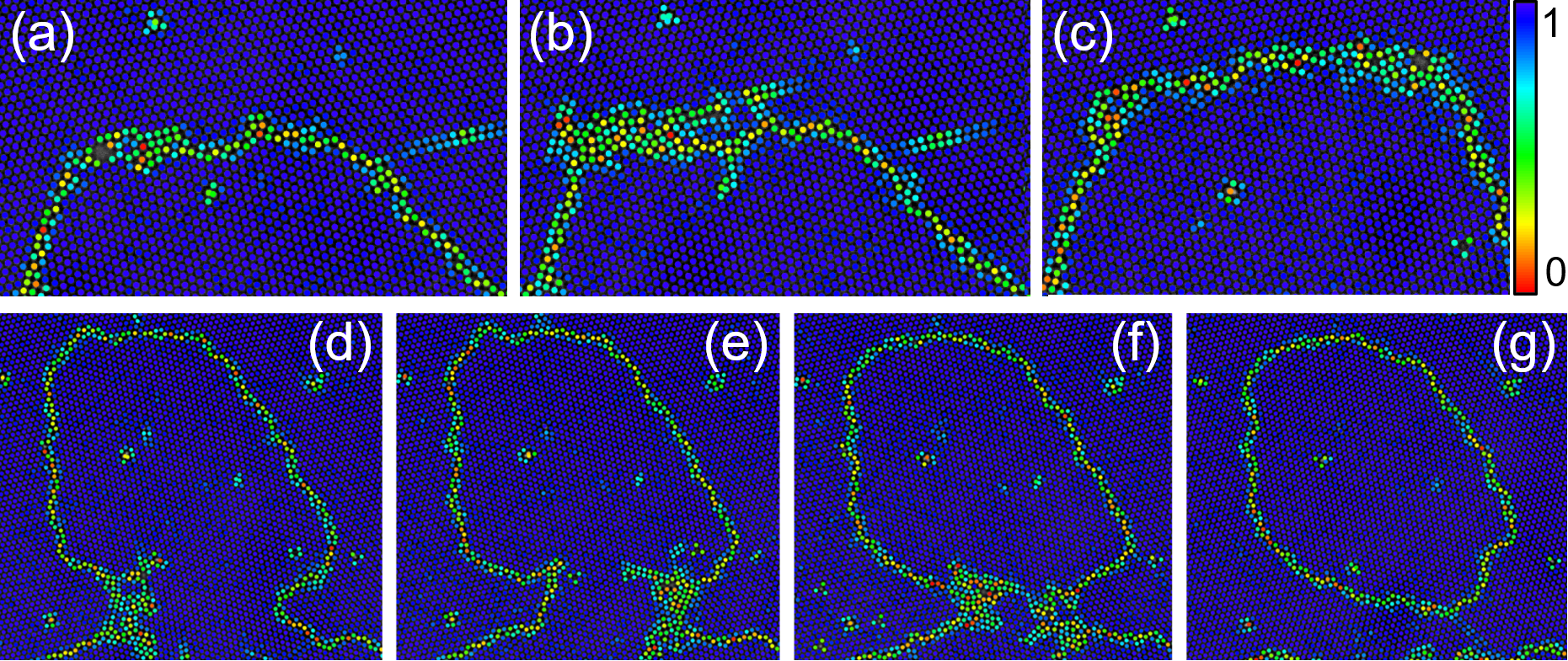}
  \caption{Sculpting an artificial grain. (a-c) We continuously move the laser to melt a band of particles directly above a grain boundary, causing the lower crystal grain to protrude into the upper crystal. (d-g) By melting the interior of the base of the protrusion, we draw the grain boundary inward to pinch off a new grain. Particle color: magnitude of $\Psi_6$.}
  \label{fig:Sculpt} 
\end{figure}

This attraction to the melted region also drives the more dramatic grain boundary manipulations we use to sculpt artificial grains. A detailed description of the procedure we employ to fabricate grains, as well as videos of the process, are provided in Supplemental Material. Briefly, to create a new custom-shaped grain, we use local melting to move a grain boundary in three stages. First, we grow a large protrusion from an existing grain (lower in Fig.~\ref{fig:Sculpt} (a-c)) by continuously moving the laser back and forth parallel to the grain boundary without completely forming a hole. This locally melts a band of particles which then recrystallizes, pulling the grain boundary upward; repeating this process causes one crystal to advance into its neighboring crystal grain (Fig.~\ref{fig:Sculpt} (a-c)). In the second stage, we disconnect the large protrusion from the original grain by melting the region in the interior of the base of the protrusion to pull the grain boundary inward, forming a neck. By repeatedly melting the interior of the neck, the two grains are eventually separated (Fig.~\ref{fig:Sculpt} (d-g)). Finally, in stage three, we further shape as desired using local melting --- the attraction to local melting is quite robust and does not depend on the details of the initial grain boundary, such as its shape or misorientation angle. Artificial grains created using this method have lifetimes ranging from hours to days. 

In conclusion, we have introduced optical blasting, a new experimental technique for perturbing 2D colloidal polycrystals. We have discovered that local melting induced by optical blasting attracts nearby grain boundaries and ultimately can be harnessed to create artificial grains of arbitrary shape. This technique can be used for new experimental investigations of colloidal grain boundaries. Optical blasting provides the ability to perturb a grain boundary to measure physical properties such as mobility and stiffness, which determine the dynamics of grain boundary migration. Such studies could explore the open questions of how and why these depend sensitively on the presence of impurities and whether the out of equilibrium values that are commonly measured in solid state crystals are equivalent to the values measured from equilibrium thermal fluctuations ~\cite{Skinner2010, Upmanyu1999, Humphreys2004,Gokhale2013,Hu2015}. Furthermore, the unprecedented ability to create grains of arbitrary shape opens up an exciting new direction for colloidal experiments. Previously, holographic optical tweezer arrays have been used to nudge grain boundary segments, and to create circular grains in 2D colloidal crystals \cite{Irvine2013}, but such techniques can only generate a limited set of grain shapes. The creation of custom grains enables investigations of phenomena ranging from dislocation pileup at grain boundaries to grain coalescence in polycrystals.

Finally, our results suggest that local melting, caused by a heat source such as a focused ion beam, could be used to deform and sculpt grains in solid state crystals of atoms. In contrast to existing ``outside in'' techniques like work hardening, such local grain-shaping could enable internal tuning of material properties, ushering in a new paradigm for materials processing.

This work was funded by the Research Corporation for Science Advancement through a Cottrell Scholar award to SJG. We thank Elizabeth Connolly and Paul Jerger for assistance with the experimental setup.

\bibliography{Cash_LJ15541_Bib}
\end{document}